# Design of a magnetically actuated flapping wing contrivance

Jason Paul, Alex Fisher and Sridhar Ravi*
*Royal Melbourne Institute of Technology*

*sridhar.ravi@rmit.edu.au

*Abstract*— **Unmanned micro aerial vehicle research is an active area of development due to the vast potential applications. Prior work towards realizing flapping flight has achieved some success; however they have also relied heavily upon the use of rotary electric motors. These require hinges, sliders, and gears to convert rotary motion to linear reciprocating motion required for flapping flight. This approach is mechanically complex, prone to jamming, inefficient and does not scale well. Therefore, a design for a novel actuator is introduced that enables the substitution of rotary electric motors for one that generates linear reciprocating motion directly. By eliminating the need to convert between rotary and reciprocal motion, substantial efficiency and reliability improvements are possible. Further improvements to the linear actuator are also described that enable independent wing actuation without added complexity or weight. Through bilateral control, a firm command over the micro aerial vehicle's flapping dynamics and consequent flight path is possible even while passive wing rotation is still employed.**

## I. BACKGROUND

Flapping wings are the universal method for birds and flying insects to generate lift and thrust for flight.
To produce flapping motion, insect musculature has evolved two primary solutions: synchronous muscles, found in ancient insects such as dragonflies, and asynchronous muscles, found in more recently evolved insects.

Synchronous muscles contract once for every nerve impulse and operate in pairs. One set of muscles contract to raise the wing during the upstroke, after which the other set contracts to lower the wing during the downstroke.

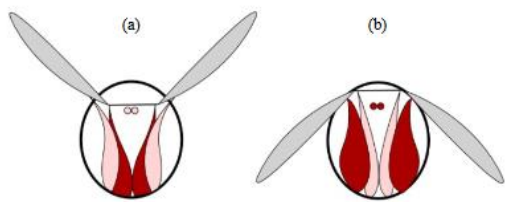

Insects using synchronous muscles typically flap at low frequencies and utilise larger wings. A locust's wings, for example, flaps at just 16 Hertz.

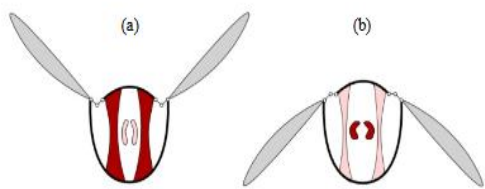

Conversely, asynchronous muscles do not directly manipulate the wings, but instead excite the insect body through inducing a resonant mechanical load which drives the wing through elastic vibration. Acting as an oscillator, the muscles go through several stroke cycles for a single nerve impulse and as a result some insects can attain flapping frequencies as high as 1000 Hertz. [1]

To achieve hovering flight the wing is pronated such that it presents a positive angle of attack during the downstroke. As the stroke terminates and motion reverses, the wing is allowed to rotate such that it also forms a positive angle of attack during the upstroke. In this manner, lift is produced during both up and down strokes, not unlike a person treading water [2].

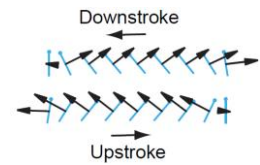

At low frequencies this can be achieved by adjusting the wing angle of attack with muscles. At high frequencies found in many insects this rotation usually cannot be directly controlled due to the flapping frequency exceeding the maximum stimulation rate of the muscles. Instead, this rotation is induced by aerodynamic torque from stroke reversal and controlled by adjusting the wing root shoulder stiffness.

To transition between hovering flight and forward flight, the majority of birds, insects, and bats tilt the long axis of the body towards the vertical such that the wings beat in the horizontal plane [3]

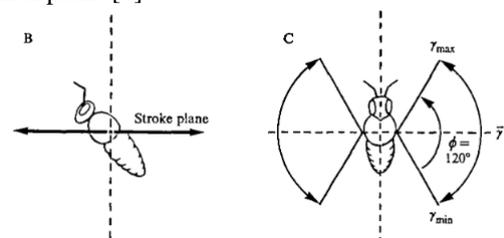

Several designs have been produced in recent years to mimic the flight of birds and insects. Most examples utilise the cheap and technologically mature rotating DC motor of which the DelFly series, developed by TU Delft University, is a successful example. To produce flapping motion, the DelFly drives the wings using a rotating DC motor without elastic components. Through gearing and mechanical components, the motor torque is transformed to flapping motion.

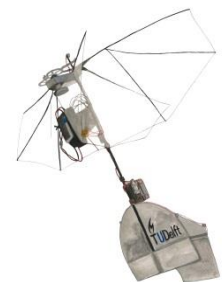

The vehicle is able to generate enough lift to carry all electronics on board including a camera and autonomous flight equipment that allows the aircraft to navigate without external input for up to 9 minutes (G.C.H.E de Croon et al. 2009).

An alternative approach that avoids the use of rotating components is the Harvard University RoboBee. Taking inspiration directly from the natural world, the RoboBee drives the wings by stimulating 'artificial muscles'.

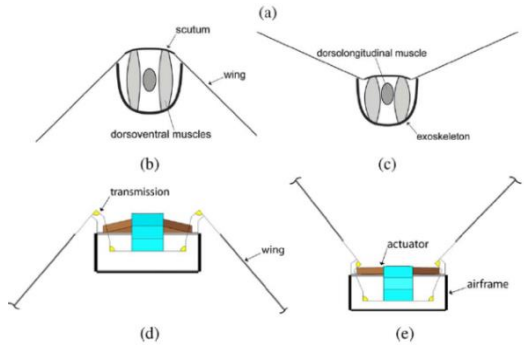

When stimulated with electricity, the piezoelectric artificial muscle deflects, driving the transmission structure and wings directly.

The RoboBee project has achieved considerable success; it is the smallest flapping vehicle yet demonstrated and first example of actively stabilised hover by independent wing actuation. While the vehicle is capable of flight, since 2008, the RoboBee relies on external power and control, which is supplied to the RoboBee via wire tether. This limitation is in part due to the high voltages required to deflect the piezoelectric material to a meaningful degree. Assuming the use of a lithium ion battery operating at 3.7 volts, the bee would require the use of a boost conversion stage with a step-up ratio of 50-100 times (Wood et al. 2005) and as of April 2018, this requirement prohibits independent flight.

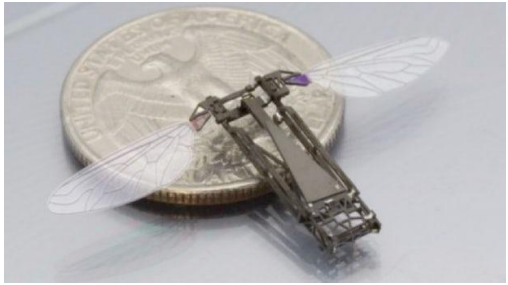

Drawing inspiration from the RoboBee in the form of the vehicle chassis and transmission structure, Zhiwei Liu et al. [4] omitted the piezoelectric muscle in favour of an oscillating cantilever beam with an attached magnet that is excited into its natural frequency by a nearby electromagnet. The working prototype weighed 0.9 grams with flapping amplitude of approximately 100 degrees and a frequency of 51.3 Hz. This was achieved at five volts which is easily attainable by current battery technology. While the vehicle was unable to achieve flight, it demonstrates that magnetically induced oscillation is a potentially feasible wing actuation method and is worthy of further development.

## II. DESIGN METHODOLOGY

The design process involves the following discrete tasks.

1. Define the system requirements.
2. Establish a theoretical model of torques opposing the motion of the wing.
3. Define the transmission geometry and determine the relationships between wing torques and actuator forces.
4. Design a magnetic actuator that is capable of producing the forces required.
5. Optimise the transmission geometry and actuator design such that sinusoidal wing flapping is produced efficiently.
6. Integrate the components into a unified structure.

## III. SYSTEM REQUIREMENTS

- 5 centimetre wing span
- 30 hertz flapping frequency.

## IV. MODELLING OF WING TORQUES

Each wing is fixed to the vehicles structure by a hinge at the wing root. By generating a moment about the wing root flapping motion is produced. From the flapping motion of the wings arises opposing torques comprised primarily from two components:

1. $\tau_{drag}$ - Torque due to aerodynamic drag which opposes the movement of the wing through the air.
2. $\tau_{inertial}$ - Torque opposing the acceleration of the wing.

### A. Torque due to drag

A reasonable estimation of the torque due to drag can be obtained by the following relation. [5]

$$\tau_{drag} = -\Omega C_d(\alpha) \cdot sign(\dot{\theta}) \cdot \omega^2 \qquad (1)$$

Where $\omega$ is the angular velocity of the wing and $\Omega$ is the drag parameter for each wing, defined as follows:

$$\Omega \equiv \frac{1}{2}\rho \int r^3 c(r) dr \qquad (2)$$

In the preceding expression, $\rho$ is the density of air and $c(r)$ is the chord dimension as a function of distance r from the wing root. The integral is evaluated from wing root to wing tip. The drag coefficient can be estimated by the following equation: [5]

$$C_D(AoA) = 1.92 - 1.55\cos(2.04 AoA - 9.82 deg) \qquad (3)$$

Where AoA denotes the angle of attack of the wing.
As the wing is assumed to stay rigid at an angle of attack of 90 degrees, this equation yields a drag coefficient of 3.46.

### B. Torque due to inertia

Torque due to inertia is defined from basic principles as the multiplication of wing angular acceleration with the mass moment of inertia about the wing root.

$$\tau_{inertial} = \alpha \cdot I_{wing} \qquad (4)$$

### C. Combined drag and inertia prediction

To obtain a reasonable estimate of the torques required to be opposed by the actuator, an example wing has been defined with the following properties. It uses a stylised cross section loosely representing an insects wing however it is composed only from triangular components to simplify the required calculations.

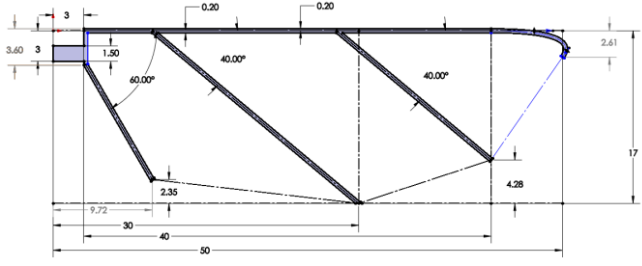

Figure 1: Hypothetical wing design

Table 1: Properties of the hypothetical wing

| | |
|---|---|
| *Total mass* ($g$) | 0.09 |
| *Mass moment of inertia about the wing root* ($g \cdot mm^2$) | 59.24 |
| *Integrated drag parameter* ($\Omega$) | $9.94 \cdot 10^{-9}$ |

Assuming the actuator is appropriately designed and can drive the wing in a sinusoidal trajectory at the targeted frequency, the wing angular position, velocity and acceleration can be described by the following equations.

$$\theta_1 = \sin(2\pi ft)\,\theta_{1_{max}} \quad (5)$$

$$\omega_C = 2\pi f \cos(2\pi ft)\,\theta_{1_{max}} \quad (6)$$

$$\alpha_C = -4\pi^2 f^2 \sin(2\pi ft)\,\theta_{1_{max}} \quad (7)$$

The torques generated by the wing are solved by applying equations 1 and 4.

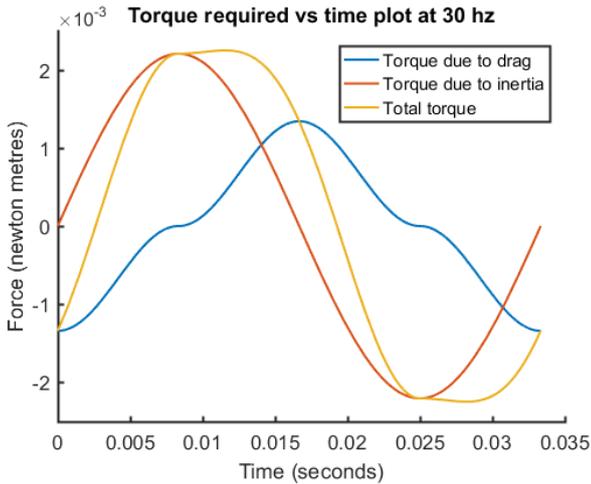

Figure 2: Torques generated by a flapping wing

The above plot has been generated for a single cycle of the wings. Of interest is the fact that at even a relatively moderate frequency of 30 hertz, the inertial torque is dominating the drag torque.

## V. TRANSMISSION DESIGN

### A. Transmission geometry

A simple crank and slider is employed to transmit the forces from the coil to the wings.

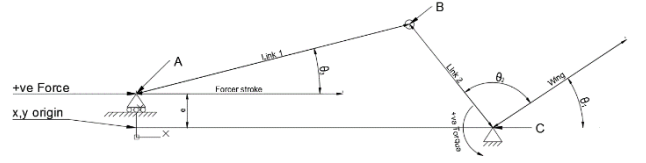

Figure 3: Transmission geometry

The coil 'forcer' is attached at point 'A' and is permitted to move horizontally a given distance (s) denoted by 'Forcer stroke'. A coil positioned at zero stroke results in maximum positive wing deflection $\theta_1$ (as illustrated above), as the forcer moves towards the right it generates a moment about 'C' driving the wing clockwise.

The angle $\theta_2$ between $L_2$ and the wing axis is fixed mechanically and is a design variable. By modifying this angle, the mechanical advantage of the force can be adjusted to bias either the upstroke or the downstroke. Additionally, offsetting the forcer stroke axis from the wing pivot point ($e$) also adjusts the mechanical advantage. By manipulating $e$ and $\theta_2$ an appropriate mechanical advantage vs stroke relationship can be established, and the non-linear force output of the coil compensated for.

### B. Force transmission properties

In reality the transmission converts linear force from the actuator to a torque about the wing pivot. In this analysis the reverse is assumed to be true, that the wing is driven in a perfect sinusoidal trajectory at the target frequency. In this way the torques resulting from the motion of the wing can be converted to the forces that would be required from the actuator to oppose them exactly and generate this motion.

From the geometry described in Figure 3, the following relationships between the wing torques and dynamics and actuator forces and dynamics are derived.

1) Determination of link lengths $L_1$ and $L_2$

$L_1$ and $L_2$ are defined as the magnitude of the vectors $\overrightarrow{AB}$ and $\overrightarrow{BC}$ respectively.

$L_2$ can be solved directly by the following equation:

$$L_2 = \frac{s(s-2d)}{2\{\sin(\theta_{1_{max}})[\sin(\theta_2)(s-2d) - 2e \cdot \cos(\theta_2)] + s \cdot \cos(\theta_{1_{max}})\cos(\theta_2)\}} \quad (8)$$

Substituting $L_2$ into the following expression will yield link length $L_1$

$$L_1 = \sqrt{\left(d + \cos(-\theta_{1_{max}} + \theta_2)L_2 - s\right)^2 + \left(\sin(-\theta_{1_{max}} + \theta_2)L_2 - e\right)^2} \quad (9)$$

2) Forcer position as a function of wing angle

With link lengths known, forcer position as a function of wing angle can be calculated via trigonometry.

$$A_x = L_2 \cos(\theta_2 + \theta_1) - L_1 \cos(\theta_3) \quad (10)$$

Where $\theta_3$ is given by the following:

$$\theta_3 = \sin^{-1}\frac{\sin(\theta_1 + \theta_2)L_2 - e}{L_1} \quad (11)$$

3) Force from torque

A torque about $C$ will produce a force
At A that is equal to:

$$F_{\overline{BA}} = \frac{-\tau \cos(\theta_3)}{L_2 \sin(\theta_1 + \theta_2 - \theta_3)} \quad (12)$$

4) Forcer acceleration from wing dynamics

The angular acceleration of the wing is linked to the linear acceleration of the forcer by the following relationship:

$$\ddot{A}_x = \frac{\frac{L_2 \alpha_C}{\csc(\theta_3 - \theta_1 - \theta_2)} + L_1 \omega_A^2 + L_2 \omega_C^2 \cos(\theta_3 - \theta_1 - \theta_2)}{\cos(\theta_3)} \quad (13)$$

Where $\alpha_C$ and $\omega_C$ are the angular acceleration and angular velocity about the wing root (point C) respectively. $\omega_A$ Is the angular velocity of link $L_1$ and is given by:

$$\omega_A = -\frac{\omega_C L_2 \cos(\theta_1 + \theta_2)}{L_1 \cos(\theta_3)} \quad (14)$$

Where $\omega_C$ is defined by equation 6.

5) Mechanical advantage

Mechanical advantage is the ratio between the component of force applied at B perpendicular to $\overrightarrow{BC}$ and the force applied at A and is given by the following equation.

$$M.A = \frac{\sin(\theta_1 + \theta_2 - \theta_3)}{\cos(\theta_3)} \quad (15)$$

For the case where the stroke axis offset ($e$) is non zero and angle $\theta_2$ is not 90 degrees, (see Figure 3) the mechanical advantage of the transmission will bias either the upstroke or the downstroke.

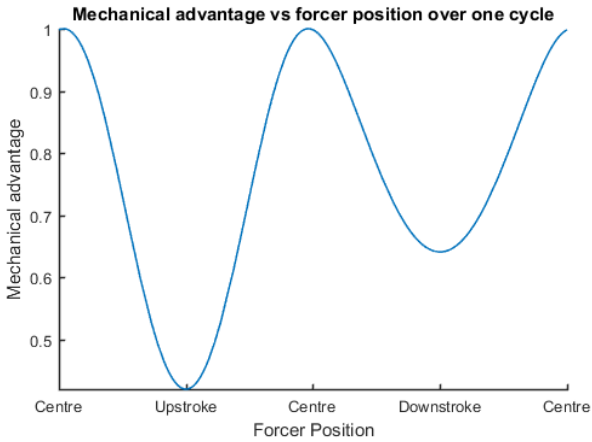

Figure 4: Mechanical advantage vs stroke position

The derivations of these equations can be found in appendix A.

*C. Designing in elasticity*

Referring again to Figure 2, it is clear that the inertial torques generated by the wing exceed the torques due to drag. While the wing mass is small, when operating at 30 hertz, the acceleration it undergoes is considerable.

Therefore, the introduction of a spring has the potential to drastically reduce the work needed to be done by the forcer. The spring will serve to decelerate the wing at the termination of each stroke, before propelling the wing back in the reverse direction. If designed with the correct stiffness, the spring can completely negate these inertial loads, leaving the forcer free to only counter the drag torque.

For the purposes of this design process, the spring is assumed to operate on the forcer as a linear spring rather than as a torque spring acting on the wing hinge. This is due to envisaged difficulty of constructing and assembling a torque spring of specific stiffness and small enough to fit the wing hinge.

Inertial forces at the forcer are primarily due to two factors, the inertial torque of the wing transmitted through the transmission, and the inertia of the forcer itself.

$$F_{Wing\,Inertia} = \frac{-\tau_{inertia} \cos(\theta_3)}{L_2 \sin(\theta_1 + \theta_2 - \theta_3)} \quad (16)$$

$$F_{Forcer\,inertia} = \ddot{A}_x \cdot m_{Forcer} \quad (17)$$

And the total force due to inertia is simply the sum of the above.

$$F_{Total\,inertia} = F_{Wing\,Inertia} + F_{Forcer\,inertia} \quad (18)$$

While the inertial torques generated by the wing are equal for both the up and down strokes, for cases where the transmission mechanical advantage does not equally bias the upstroke or downstroke, the force transmitted to the forcer may differ between strokes.
Furthermore, the acceleration of the forcer ($\ddot{A}_x$, given by equation 13) is not sinusoidal due to the geometry of the transmission and the acceleration during the upstroke differs from that during the downstroke.

Therefore, for a spring to counter the inertial loads equally throughout the stroke it will need to be offset such that a larger spring force is provided when mechanical advantage is low, and vice versa. This offset can be calculated using the following equation.

$$offset = \frac{s}{2}\left[\frac{1-R}{1+R}\right] \quad (19)$$

In the preceding equation R is the ratio of inertial loads at the termination of the upstroke and downstroke and is given by the following:

$$R = \frac{F_{Total\,inertia_{upstroke}}}{F_{Total\,inertia_{downstroke}}} \quad (20)$$

Finally, the stiffness of the spring element is given by the following equation, derived from Hook's law of springs.

$$F_{spring} = K_{spring} \cdot (A_x - \frac{s}{2} + offset) \quad (21)$$

The offset is calculated to balance the forces at both up and down stroke, therefore spring stiffness can be calculated by dividing the force due to inertia at either stroke by the forcer displacement.

At the upstroke, $A_x$ is zero (Figure 3) and so spring stiffness is given by:

$$K_{spring} = \left| \frac{F_{Total\ Inertia_{upstroke}}}{-\frac{S}{2} + offset} \right| \quad (22)$$

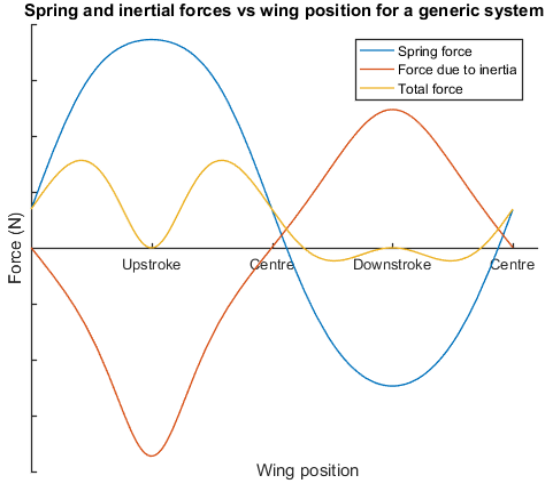

Figure 5: Springs can counter inertial loads

Because the inertial forces are warped by the transmission geometry, the linear performance of the spring is unable to compensate exactly except at the termination of the up and down stroke as was intended. It is however, at these points that the loads are the highest and as a result, from Figure 5 it can be seen that the inertial forces have been reduced substantially. Further improvement is also possible with additional optimisation of the transmission geometry.

### D. Force and power requirements

Total force required from the actuator is then the sum of force due to wing inertia, forcer inertia, spring force and wing drag. Wing inertia, forcer inertia, spring force can be found from equations 16, 17, 21 respectively, force from drag can be found from substituting equation 1 into equation 12.
Plotting the above equations into a generic transmission geometry yields the following:

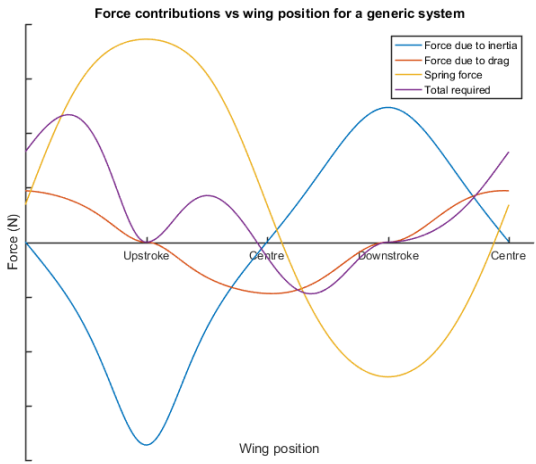

Figure 6: Required force and composition

The average power required to drive the system is now also defined, being the integral of total force with respect to displacement divided by the time required to complete a stroke.

$$Avg\ power = \frac{1}{f} \int_S^0 (F_{Drag} + F_{spring} + F_{Total\ inertia}) dA_x \quad (23)$$

## VI. ACTUATOR DESIGN

### A. The magnetic forcer

The actuator forcer is composed of a single cylindrical neodymium magnet and a coaxial copper coil with a larger inner diameter than the magnet diameter.

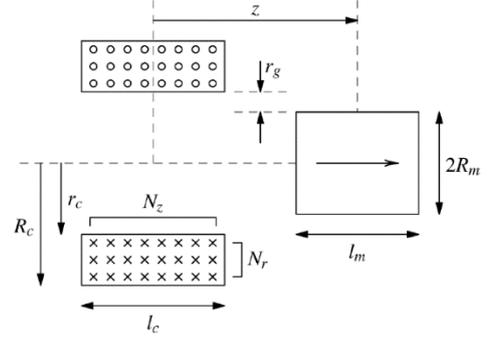

Figure 7: Coaxial magnet and coil geometry

Current flowing in a coil will produce a magnetic field directly proportional to the applied current. Placing the coil in the vicinity of another magnetic field will result in a net force given by the following equations. [2]

$$F = \frac{B_r NI}{l_c[R_c - r_c]} \int_{-l_c/2}^{l_c/2} \int_{r_c}^{R_c} \sum_{e_1}^{\{1,-1\}} [e_1 m_6 f_{z3}] dr_2 dz_2 \quad (24)$$

The sigma notation is non-standard, for the purposes of the above equation the intended interpretation is to substitute 'e1' into the equation with values -1 and then +1 and to sum the two results. The result is not zero due to e1 also being included in the 'm6' intermediate parameter seen below.

$$f_{z3} = \left[1 - \frac{1}{2} m_5\right] K(m_5) - E(m_5) \quad (25)$$

$$m_5 = \frac{4 R_M r_2}{m_6^2}, m_6^2 = [R_m + r_2]^2 + \left[z + \frac{1}{2} e_1 l_m - z_2\right]^2 \quad (26)$$

Table 2: Definition of parameters

| | |
|---|---|
| $B_r$ | Magnetic remanence of the permanent magnet |
| $N$ | Number of turns comprising the coil. |
| $I$ | Current directed through the coil. |
| $R_c$ | Outer radius of the coil. |
| $r_c$ | Inner radius of the coil. |
| $l_c$ | Length of the coil. |
| $R_m$ | Radius of the permanent magnet. |
| $l_m$ | Axial length of the permanent magnet. |
| $K(m_5)$ | Complete elliptical integral of the first kind with parameter m5. |
| $E(m_5)$ | Complete elliptical integral of the second kind with parameter m5. |

| $z_2, r_2$ | Variables of integration. |
|---|---|
| $z$ | Distance between coil and magnet centroids. |

An important note to make from equation 24 is that the force produced is directly proportional to the applied current. As a result, while there may be some optimum coil and magnet geometry, at this early stage in development it is appropriate to vary their respective parameters to satisfy structural and form factor considerations. The applied current can simply be varied to compensate.

The coil and magnet properties ultimately selected are as follows.

Table 3: Properties of the magnet and copper coil

| | |
|---|---|
| **Magnet radius:** | 4 millimetres |
| **Axial length of the magnet:** | 5 millimetres |
| **Magnetic remanence:** | 1.3 Tesla |
| **Number of turns comprising the coil:** | 30 Turns |
| **Coil current magnitude** | 2.9 Amperes |
| **Inner radius of the coil** | 5 millimetres |
| **Outer radius of the coil** | 5.6 millimetres |
| **Axial length of the coil** | 1.5 millimetres |
| **Mass of the coil** | 0.4 grams |

Given the dimensions described in Table 3, force between the coil and the magnet is solved via the numerical trapezoidal method within MATLAB yielding the following force vs displacement curve.

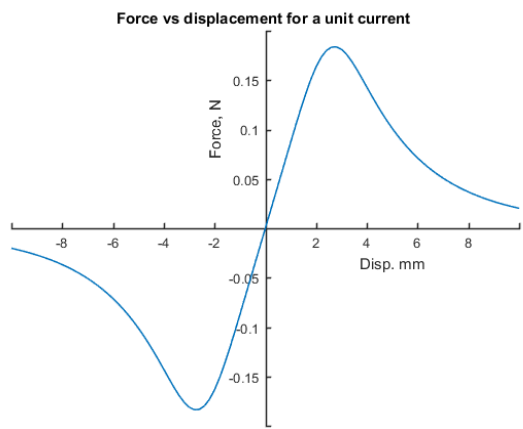

Figure 8: Magnetic force vs displacement

Force is weakest when the coil centroid and magnet centroid overlap and greatest when the coil centroid is positioned over one of the magnet faces. As distance increases further, the magnetic force continues to decay exponentially.

*B. Current waveform*

As force production of the coil is directly proportional to applied current, the current waveform required can be obtained by dividing the required force (Figure 6) by the magnetic force distribution for a unit current (Figure 8).

Performing this calculation for the generic case yields the following:

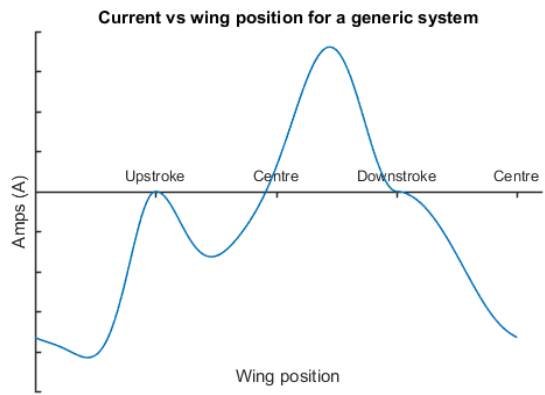

Figure 9: Current waveform to produce required force

The current waveform is highly irregular, this is to be expected as it reflects the irregular nature of the forces required to offset the nonlinear magnetic force production, the nonlinear force magnification of the transmission and the sinusoidal end actuation of the wing. Plotting the final distribution of forces yields the following.

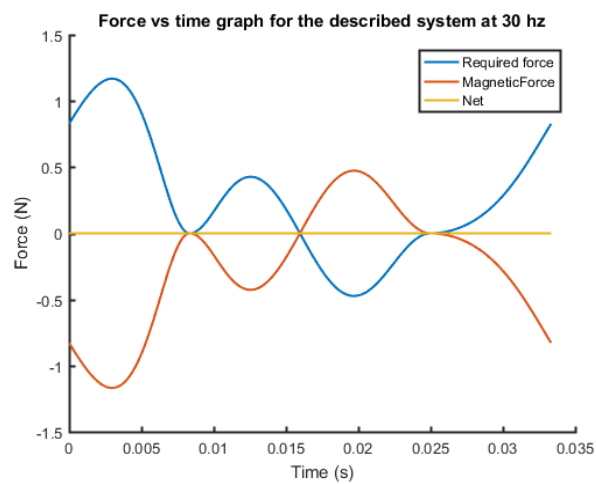

Figure 10: Magnetic force vs required force.

As expected, the actuator is not doing any work during the termination of each stroke (denoted by up and downstroke) as the spring is absorbing all the inertial loads and then returning it to the wing. During the mid-stroke (denoted by centre) the actuator is working the hardest, overcoming drag as designed. If this force distribution can be produced in reality, extremely efficient wing actuation can be expected.

VII. COMPONENT SIZING AND OPTIMISATION

As the magnetic and coil geometries were selected previously, the transmission geometry will be optimised around these choices to arrive at a final design solution.

*A. Stroke length and separation distance (d) determination*

Referring to Figure 8, it can be seen that magnetic force peaks at a displacement of approximately 3.2 mm. It would be advantageous for the displacement to be as close as possible to this value as possible during the mid-stroke where aerodynamic forces are highest (Figure 2).
The spring will absorb the inertial loads at the termination of each stroke (Figure 5) and there is therefore no need for high force production from the coil at these times. As a result the displacement can be off nominal.

Therefore a 5 millimetre stroke beginning at 1.6 millimetres and terminating at 6.6 millimetres is chosen to approximately meet these needs.

For the separation distance (d) between point A and point C in Figure 3, a value of 10 millimetres is selected. This is to ensure a reasonable length for $L_1$ and $L_2$ to facilitate ease of manufacture.

*B. Transmission geometry optimisation*

The 2 design parameters that are to be investigated is the angle $\theta_2$ between $L_2$ and the wing axis and the forcer stroke axis from the wing pivot point($e$) (see Figure 3 for definitions). A parametric study varying the above design parameters was conducted for a flapping frequency of 30 hertz, repeating the process used to obtain the total force, average power and current waveform for each permutation. The results are plotted below.

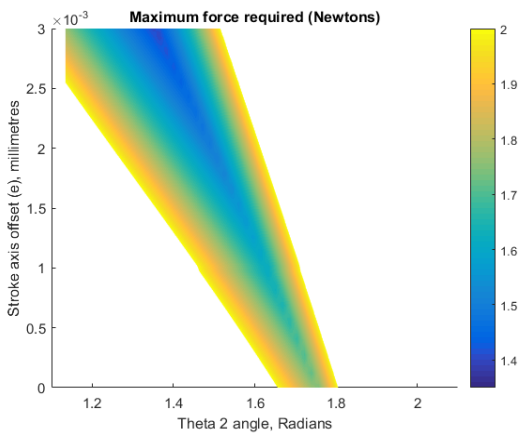

Figure 11: Maximum force required from the actuator

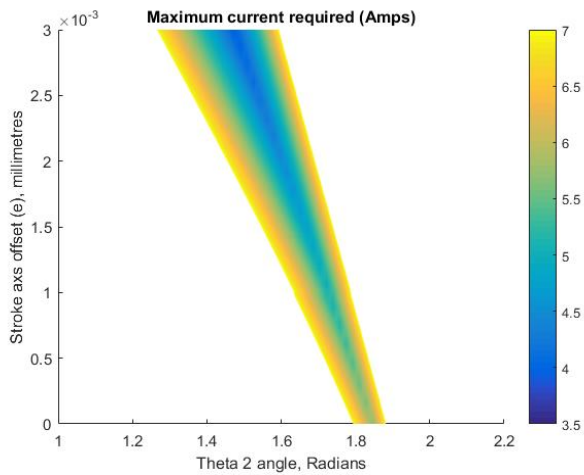

Figure 12: Maximum current required.

From Figure 11 and Figure 12, it's clear that there is a linear relationship between the two design parameters and the optimum force and current properties. This relationship has not been investigated here but will hopefully be revealed in future work.
Outside of this linear region (the blue stripes), the force and current values required accelerate exponentially. The results have been truncated therefore, to limit the design space to reasonable values. The white areas are the discarded values

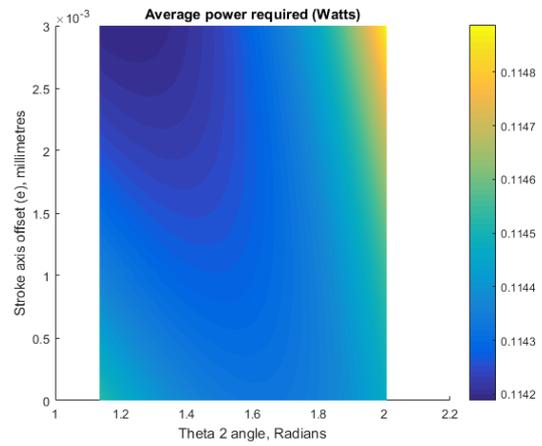

Figure 13: Average power required from the actuator

The average power required varies only slightly about 0.11 watts as it is primarily affected by the wing planform and flapping frequency – which are both held constant in this study.

Comparing Figure 11 and Figure 12 again, it can be seen that as the design parameters vary from the ideal, the force required grows more slowly than the required current. This implies that design decisions should be made primarily based on required current – which has battery, electronics and heating considerations.

Finally, the best combination explored for the maximum required current is at 3mm stroke axis offset (e), and at a $\theta_2$ of 1.477 radians, this results in a maximum required current of 3.94 amps. 1.477 radians (84.8 degrees) is close to a right angle, for the right angle case the maximum required current becomes 4.35 amps at a stroke axis offset (e) of 2.3 millimetres. A 10% gain in maximum current is deemed an acceptable sacrifice to significantly simplify manufacture.

Below the selected parameters are summarised, followed by the calculated parameters that together complete the system.

Table 4: Selected parameters

| | |
|---|---|
| **Stroke axis offset** (e) | 2.3 millimetres |
| **Angle $\theta_2$** | 90 degrees |
| **Separation distance** (d) | 10 millimetres |
| **Stroke length (s)** | 5 millimetres |

Table 5: Calculated component sizes and system properties

| | |
|---|---|
| **Link $L_1$ length** | 7.5 millimetres |
| **Link $L_2$ length** | 2.9 millimetres |
| **Spring offset** | 0.48 millimetres |
| **Spring stiffness** | 644.6 Newton metres |
| **Average power required** | 0.11 Watts |
| **Maximum current required** | 4.35 Amperes |
| **Maximum force required** | 1.9 Newtons |

## VIII. FINAL DESIGN

*A. Independent wing actuation*

Until this point all modelling has concerned the actuation of a single wing individually. This allows the possibility of duplicating the transmission and actuator for the other wing. By enabling independent wing actuation in this way, full bilateral control of the wings is possible. Complete duplication is a waste however, and the challenge of this approach is to arrive at a design solution that maintains wing independence while allowing the actuator to share as many components as possible between wings.

One such component is the magnet, referring to Figure 8 it can be seen that magnetic force is reflected about the Y axis. This means that it's possible for two coils to share the same magnet and produce the same forces (albeit reversed) provided they do not Collide or Interfere with each other's magnetic fields.

As coil axial length is 1.5 millimetres and the minimum stroke length was designed to be 1.6 millimetres between the coil and magnet centroids, there is no risk of collision. Since the coils are the closest at minimum stroke, if there is any risk of magnetic interference it would be at this point. However, since minimum forcer stoke corresponds with the maximum wing upstroke, total force required is zero (Figure 6) therefore the current applied to the coil is zero (Figure 9) and, as a result, the magnetic force produced is zero (Figure 10). Because the coils have zero current when they are at their closest point, they cannot interfere.

*B. Structural solution to elasticity*

Drawing inspiration from the natural world, the intention is to find a structural solution to elasticity. The inbuilt flexibility of the flapping vehicles 'thorax' should provide the restoring force required to oppose the wing inertial forces.

Following prior work [4], the first attempt at this concept was a simple cantilever oscillator.

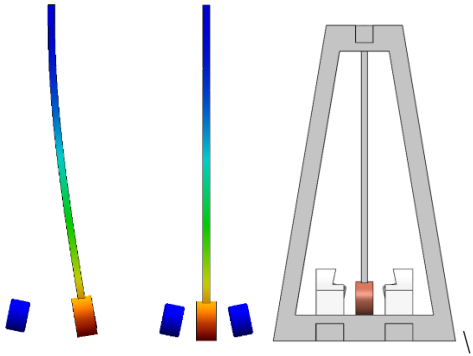

The cantilever proved successful, and validated the magnetic force and vibration models thus far developed. Target amplitude oscillation was achieved at 30 hertz using only 32 milliwatts, suggesting the concept could realistically be applied to flapping wing applications.

However, the length of the cantilever becomes significant when designing small scale actuators. This is especially true when considering the size of the chassis structure required to fix the cantilever firmly in place.
A concept devised to avoid this issue draws its inspiration from the bow and arrow. By curving the cantilever into a parabola, the actual length of the elastic structure can be quite long and yet still only occupy a small footprint.

Finally, by closing the bow structure into a hoop, tip rotation is eliminated, and the coil is free to oscillate linearly along the stroke axis leading to the final configuration displayed below.

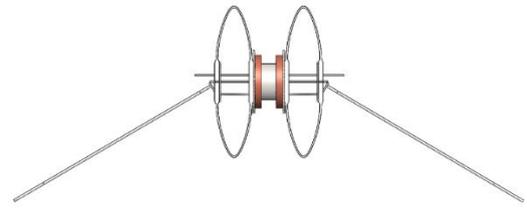

Figure 14: Front view, upstroke

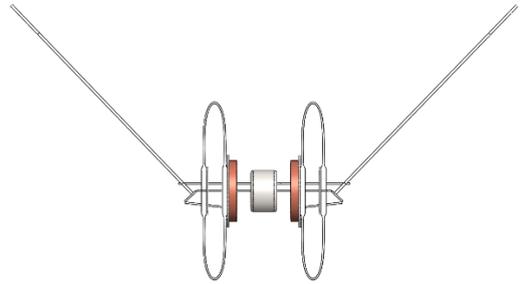

Figure 15: Front view, downstroke

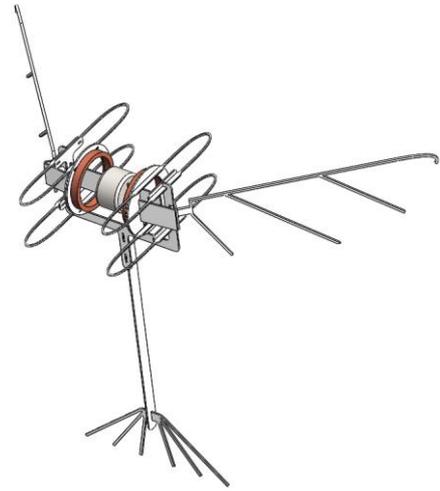

Figure 16: Perspective view, downstroke

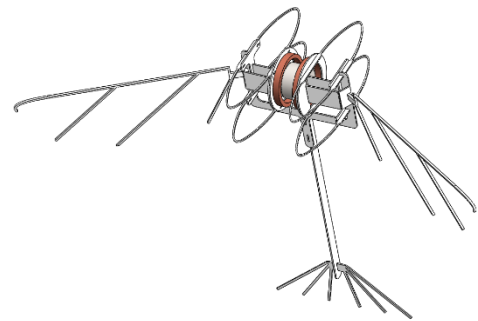

Figure 17: Perspective view, upstroke

IX. RESULTS AND CONCLUSION

Manufacturing is due to be completed in the coming days (25/may/2018), once the model has been tested this paper will be updated with the final results.

## APPENDIX A: DERIVATION OF EQUATIONS

### A. Sign conventions

1. Counter clockwise angles, angular velocities and angular accelerations are positive. Clockwise is negative.
2. Positive x is to the right.
3. Positive y is vertical upwards.

The following relationships are therefore inferred:

1. Positive torque applied at the wing root produces positive angular acceleration and vice versa.

2. Positive force applied at the forcer produces negative torque at the wing root and vice versa.

### B. Determination of link lengths L1 and L2

First, describe link length $L_1$ entirely in terms of known quantities and $L_2$ then solve simultaneously.

$$L_1 = \sqrt{(B_x - A_x)^2 + (B_y - c)^2}$$

Where:
$B_x = d + \cos(\theta_1 + \theta_2) L_2$ and, $B_y = \sin(\theta_1 + \theta_2) L_2$

Conditions at upstroke:
$A_x = 0$
$\theta_1 = 1 \cdot \theta_{1_{max}} = \theta_{1_{max}}$

Conditions at downstroke:
$A_x = s$
$\theta_1 = -1 \cdot \theta_{1_{max}} = -\theta_{1_{max}}$

Two equations are therefore:

$$L_{1_{downstroke}} = \sqrt{(d + \cos(-\theta_{1_{max}} + \theta_2) L_2 - s)^2 + (\sin(-\theta_{1_{max}} + \theta_2) L_2 - c)^2}$$

$$L_{1_{upstroke}} = \sqrt{(d + \cos(\theta_{1_{max}} + \theta_2) L_2)^2 + (\sin(\theta_{1_{max}} + \theta_2) L_2 - c)^2}$$

Since the link lengths do not change, $L_{1_{upstroke}}$ and $L_{1_{downstroke}}$ are equated and solved for $L_2$ yielding:

$$\boxed{L_2 = \frac{s(s - 2d)}{2\{\sin(\theta_{1_{max}})[\sin(\theta_2)(s - 2d) - 2c \cdot \cos(\theta_2)] + s \cdot \cos(\theta_{1_{max}})\cos(\theta_2)\}}}$$

$L_1$ can then be solved directly via substitution.

### C. Force from torque

A torque ($\tau$) about $C$ will produce a force at $B$ acting perpendicular to the vector $\overrightarrow{CB}$ with a magnitude equal to $\frac{\tau}{L_2}$.
The force directed down $\overrightarrow{BA}$ then is equal to:

$$F_{\overrightarrow{BA}} = \frac{-\tau}{L_2 \sin(\theta_1 + \theta_2 - \theta_3)}$$

And the force generated at A equal to:
$$F_{A_x} = \cos(\theta_3) F_{\overrightarrow{BA}}$$
$$F_{A_y} = \sin(\theta_3) F_{\overrightarrow{BA}}$$

Simplified:

$$\boxed{F_{A_x} = \frac{\cos(\theta_3) - \tau}{L_2 \sin(\theta_1 + \theta_2 - \theta_3)}}$$

### D. Forcer acceleration from wing dynamics

First solve for acceleration at B with respect to the wing and forcer separately:
$$\ddot{B} = \ddot{C} + \alpha_C \times \overrightarrow{BC} + \omega_C^2 \times \overrightarrow{BC}$$
$$\ddot{B} = \ddot{A} + \alpha_A \times \overrightarrow{BA} + \omega_A^2 \times \overrightarrow{BA}$$

Point C is fixed, therefore the acceleration of C ($\ddot{C}$) is zero and it is discarded.

Evaluating cross products:
$$\ddot{B} = [-\alpha_C L_2 \sin(\theta_1 + \theta_2) - \omega_C^2 L_2 \cos(\theta_1 + \theta_2)]\hat{\imath} + [\alpha_C L_2 \cos(\theta_1 + \theta_2) - \omega_C^2 L_2 \sin(\theta_1 + \theta_2)]\hat{\jmath}$$

$$\ddot{B} = \ddot{A}_{\hat{\imath}} + [-\alpha_A L_1 \sin(\theta_3) - \omega_A^2 L_1 \cos(\theta_3)]\hat{\imath} + [\alpha_A L_1 \cos(\theta_3) - \omega_A^2 L_1 \sin(\theta_3)]\hat{\jmath}$$

Collecting like terms:
$$\ddot{B}_{\hat{\imath}} = \ddot{A}_{\hat{\imath}} - \alpha_A L_1 \sin(\theta_3) - \omega_A^2 L_1 \cos(\theta_3)$$
$$= -\alpha_C L_2 \sin(\theta_1 + \theta_2) - \omega_C^2 L_2 \cos(\theta_1 + \theta_2)$$

$$\ddot{B}_{\hat{\jmath}} = \alpha_A L_1 \cos(\theta_3) - \omega_A^2 L_1 \sin(\theta_3)$$
$$= \alpha_C L_2 \cos(\theta_1 + \theta_2) - \omega_C^2 L_2 \sin(\theta_1 + \theta_2)$$

Utilising the fact the vertical component of $a_A$ is zero:
$$\ddot{A}_{\hat{\jmath}} = 0 = \alpha_C L_2 \cos(\theta_1 + \theta_2) - \omega_C^2 L_2 \sin(\theta_1 + \theta_2) - \alpha_A L_1 \cos(\theta_3) + \omega_A^2 L_1 \sin(\theta_3)$$

$$\alpha_A = \frac{\alpha_C L_2 \cos(\theta_1 + \theta_2) - \omega_C^2 L_2 \sin(\theta_1 + \theta_2) + \omega_A^2 L_1 \sin(\theta_3)}{L_1 \cos(\theta_3)}$$

Subbing in $\alpha_A$ to $a_{b_i}$ to solve for angular acceleration at the wing and simplifying yields;

$$\alpha_C = \frac{\csc(\theta_3 - \theta_1 - \theta_2)[-L_1 \omega_A^2 + L_2 \omega_C^2 \cos(\theta_3 - \theta_1 - \theta_2) + \ddot{A}_i \cdot \cos(\theta_3)]}{L_2}$$

Rearranging and solving for $\ddot{A}_i$, (or $\ddot{A}_x$ in x, y) coordinates, yields the final equation:

$$\boxed{\ddot{A}_x = \frac{\frac{L_2 \alpha_C}{\csc(\theta_3 - \theta_1 - \theta_2)} + L_1 \omega_A^2 + L_2 \omega_C^2 \cos(\theta_3 - \theta_1 - \theta_2)}{\cos(\theta_3)}}$$

The above equation relies on the angular velocity of link $L_1$ at A ($\omega_A$), it is advantageous to recast this in terms of $\omega_C$ which is a defined property (equation 6). The process to find $\omega_A$ in terms of $\omega_C$ is as follows:

First solve for velocity at B with respect to the wing and forcer separately:
$$\dot{B} = \omega_C \times \overrightarrow{BC}$$
$$\dot{B} = \dot{A}_i + \omega_A \times \overrightarrow{BA}$$

$$\dot{A}_i = \omega_C \times \overrightarrow{BC} - \omega_A \times \overrightarrow{BA}$$

Evaluating cross products yields the following
$$\dot{B} = -\omega_C L_2 \sin(\theta_1 + \theta_2)\hat{\imath} + \omega_C L_2 \cos(\theta_1 + \theta_2)\hat{\jmath}$$
$$\dot{B} = \dot{A}_i - \omega_A L_1 \sin(\theta_3)\hat{\imath} + \omega_A L_1 \cos(\theta_3)\hat{\jmath}$$

Equating terms and solving for $\dot{A}$ :
$$\dot{A} = -\omega_C L_2 \sin(\theta_1 + \theta_2)\hat{\imath} + \omega_C L_2 \cos(\theta_1 + \theta_2)\hat{\jmath}$$
$$- \omega_A L_1 \sin(\theta_3)\hat{\imath} + \omega_A L_1 \cos(\theta_3)\hat{\jmath}$$

Utilising the fact that the vertical component of $\dot{A}$ is zero:
$$\dot{A}_{\hat{\jmath}} = 0 = -\omega_C L_2 \cos(\theta_1 + \theta_2) - \omega_A L_1 \cos(\theta_3)$$

$$\boxed{\omega_A = -\frac{\omega_C L_2 \cos(\theta_1 + \theta_2)}{L_1 \cos(\theta_3)}}$$

### E. Mechanical advantage

Equal to the component of force applied at B perpendicular to $\overrightarrow{BC}$, divided by the force applied at A.

$$F_{B_{Perpendicular}} = \frac{\tau}{L_2}$$

$$F_A = \frac{-\cos(\theta_3)\tau}{L_2 \sin(\theta_1 + \theta_2 - \theta_3)}$$

$$M.A = \frac{F_B}{F_A} = \frac{\tau}{L_2} \cdot \frac{L_2 \sin(\theta_1 + \theta_2 - \theta_3)}{-\cos(\theta_3)\tau}$$

$$\boxed{M.A = -\frac{\sin(\theta_1 + \theta_2 - \theta_3)}{\cos(\theta_3)}}$$